\documentclass[ba,usenames,dvipsnames]{imsart}
\pubyear{2023}
\volume{TBA}
\issue{TBA}
\firstpage{1}
\lastpage{1}

\usepackage{amssymb}
\usepackage{subcaption}
\usepackage{amsthm}
\usepackage{amsmath}
\usepackage{natbib}
\usepackage[colorlinks,citecolor=blue,urlcolor=blue,filecolor=blue,backref=page]{hyperref}
\usepackage{graphicx}
\usepackage{algorithm2e}
\usepackage{fancyvrb}
\VerbatimFootnotes
\usepackage{placeins}

\startlocaldefs

    \theoremstyle{plain}
    \newtheorem{theorem}{Theorem}
    \newtheorem{proposition}[theorem]{Proposition}
    \theoremstyle{definition}
    \newtheorem{example}{Example}
    
    \newcommand{\E}{\mathbb{E}}
    \newcommand{\cor}{\mbox{Corr}}
    \def\*#1{\boldsymbol{#1}}
    \def\~#1{{\cal #1}}

    \newcommand{\green}[1]{{\color{black} #1}}
    
\endlocaldefs

\begin{document}


\begin{frontmatter}
\title{Posterior shrinkage towards linear subspaces}
\runtitle{Posterior shrinkage towards linear subspaces}

\begin{aug}
\author{\fnms{Daniel K.} \snm{Sewell} \thanksref{addr1} \ead[label=e1]{daniel-sewell@uiowa.edu}}

\runauthor{D. K. Sewell}

\address[addr1]{
University of Iowa, 145 N. Riverside Dr., Iowa City, IA 52241 \printead{e1}
}


\end{aug}

\begin{abstract}
It is common to hold prior beliefs that are not characterized by points in the parameter space but instead are relational in nature and can be described by a linear subspace. While some previous work has been done to account for such prior beliefs, the focus has primarily been on point estimators within a regression framework.  We argue, however, that prior beliefs about parameters ought to be encoded into the prior distribution rather than in the formation of a point estimator. In this way, the prior beliefs help shape \textit{all} inference.  Through exponential tilting, we propose a fully generalizable method of taking existing prior information from, e.g., a pilot study, and combining it with additional prior beliefs represented by parameters lying on a linear subspace. We provide computationally efficient algorithms for posterior inference that, once inference is made using a non-tilted prior, does not depend on the sample size.  We illustrate our proposed approach on an antihypertensive clinical trial dataset where we shrink towards a power law dose-response relationship, and on monthly influenza and pneumonia data where we shrink moving average lag parameters towards smoothness.  Software to implement the proposed approach is provided in the R package \verb+SUBSET+ available on GitHub.
\end{abstract}


\end{frontmatter}



\section{Overview}
Prior beliefs are often reflected in Bayesian analyses by shrinking estimates towards some point $\*\theta_0$ of the parameter space $\Theta$, such as those priors inducing sparsity \citep[e.g.,][]{Park2008bayesian,carvalho2010horseshoe}.  However, at other times prior knowledge leads to beliefs that are not characterized by points in $\Theta$ but rather are relational in nature:  In a regression framework we may believe a priori that the coefficients of a polychotomous ordinal factor covariate might reflect a linear shape; In a two sample binomial context, the response rates of the two corresponding populations may be believed to be near equal; When conducting an ANOVA we may believe a priori that the multiple populations have near homoscedastic responses. The prior beliefs in these examples and in other similar situations can be encoded by shrinking our estimates towards the intersection of the parameter space and a linear subspace.

Stein-type estimators have received much attention from statisticians since the landmark paper by James and Stein \citep{james1961estimation}. Such estimators have been used repeatedly in regression settings to shrink the mean response curve towards a linear subspace contained within the span of the columns of the design matrix.  An early application of this was the work by \cite{Blaker1999class}, who developed Stein-type estimators that shrinks the mean response towards the space spanned by the principal components.
More recently, \cite{Shin2020functional} and \cite{wiemann2021adaptive} considered the context of spline regression.  By applying what they termed a functional horseshoe prior, the authors were able to shrink the mean response curve towards a linear subspace (e.g., a simple polynomial relationship between $y$ and $X$). While these results are exciting, they are limited to regression settings, only shrink the mean of the response vector, and require Gaussianity of the prior, which may be singular.
\cite{huber2021subspace} developed a somewhat similar approach for vector autoregressive models with a focus on shrinking towards factor models. \cite{An2009shrinkage} considered Stein-type estimators in the generalized linear model setting, proving some theoretical results in the frequentist paradigm; see references therein for other landmark papers on Stein-type estimators.

Deviating away from shrinking the mean response in regression settings, \cite{Oman1982shrinking} novelly focused on the regression coefficients directly.  Using an empirical Bayesian approach, Oman developed a point estimator for the regression coefficients that provided shrinkage of the least squares estimate towards its projection onto the linear subspace of interest. \cite{Lee1994shrinking} proposed a subspace ridge estimator for the regression coefficients, a generalized ridge estimator shrinking the estimates towards a linear subspace; \citeauthor{Lee1994shrinking} also showed how their point estimator can be derived through a three-stage hierarchical Bayesian approach with improper priors.  \cite{Hansen2016efficient} expanded beyond the scope of regression settings (although their work still applies in that context), proposing a general purpose point estimator that is a weighted average of the unconstrained maximum likelihood estimator (MLE) and the MLE restricted to the subspace of interest. Finally, \cite{floto2022exponentially} applied an exponentially tilted Gaussian prior for deep neural networks.

Prior work in this realm of shrinkage towards linear subspaces have focused on point estimation, and nearly all on the frequentist properties of the proposed estimators. However, if a researcher wants to shrink their estimates towards a linear subspace, it is because there is \textit{prior information} regarding the plausible values of the parameters.  Therefore, taking a Bayesian stance is the most sensible approach, where such prior information can naturally be incorporated into the prior belief distribution over the parameters.  This is in contrast to (1) ad hoc- even if reasonable- adjustments of frequentist point estimates, and (2) loss functions that act to shrink posterior-based point estimates towards a subspace.

The purpose of this paper is to provide the Bayesian practitioner highly generalizable, easily implemented, and computationally efficient methods to incorporate prior information which can be encoded by shrinking towards a linear subspace. Our approach applies shrinkage on any parameters of interest, unlike some prior work that focuses solely on the mean structure, and while certainly applicable to regression settings, it can be applied to any parametric setting. We emphasize that our approach \textit{incorporates} additional information into the prior, rather than \textit{replaces} other prior information with a specific prior distribution that conveniently induces shrinkage. Further, our approach can be applied to any proper prior, not just Gaussian priors. By adjusting the prior- and hence the posterior- rather than the point estimator, all posterior inferential statements account for prior information regarding the linear subspace. 

The remainder of the paper is as follows. In Section \ref{sec:exp_tilt_priors}, we discuss our proposed method to exponentially tilt an existing prior to append a priori knowledge or beliefs about unknown parameters lying in or near linear subspaces.  In Section \ref{sec:estimation} we provide computationally efficient methods of obtaining posterior inference under the tilted prior, including methods for when the linear subspace itself is not fully known. In Section \ref{sec:simstudy} we evaluate our method in a simulation study.  In Section \ref{sec:rda}, we illustrate our methods on two real datasets: a clinical trial evaluating antihypertensive drug treatment where we shrink towards a power law dose-response relationship, and on monthly influenza and pneumonia data where we shrink moving average lag parameters towards smoothness.
Section \ref{sec:conclusion} provides a brief summary and discussion.

\section{Exponentially tilted priors}
\label{sec:exp_tilt_priors}
\subsection{Introduction}
\label{subsec:subset_intro}
Consider the typical Bayesian data analysis set up:  Let $y$ denote the observed data with corresponding likelihood $\pi(y|\theta)$ parameterized by some $p$-dimensional unknown parameter vector $\theta\in\Theta\subseteq \Re^p$.  Let $\pi_0(\theta)$ denote the \textit{base prior density}, which reflects our prior beliefs about plausible values of $\theta$.  

While the practitioner is usually a practiced hand at setting up a prior distribution to reflect information on location or degree of uncertainty about the true value of $\theta$, often the information we feel more assured about is much more difficult to encode in $\pi_0$. Consider the simplest of examples: if we believe a priori that a placebo mean will be 1 and a treatment mean will be 5, then we can center our prior $\pi_0$ on $\theta = (\theta_{placebo},\theta_{trmnt})$ at $(1,5)$.  But what ought one to do if, on the other hand, we have little or no prior information on the exact values of the treatment effects but believe that the placebo will likely have some non-zero effect and that the treatment arm will have 5 times the effect?  In such a case  we believe that there ought to be some $x$ such that the true $\theta$ isn't too far from $(x,5x)$; in other words, we believe the true $\theta$ should be somewhere around the linear subspace defined by the span of $(1,5)'$. 

In the simplest cases such as the example given above, it may be tempting to consider a reparameterization (e.g., $\theta_{trmnt} = 5\theta_{placebo}$ or $\theta_{trmnt} = 5\theta_{placebo} + noise$).  However, there are distinct disadvantages to using reparameterization as a general approach to encoding relational information on the parameters.  First, because of the lack of generalizability of reparameterization, determining how to perform estimation must occur on a case-by-case basis, and in some instances estimation may prove challenging or computationally onerous. Second, it is not obvious how to integrate relational information through a reparameterization with other non-relational prior information, such as that obtained from a pilot study or literature review.  Third, it may not be possible to perform such a reparameterization when the parameter space is bounded.  Fourth, reparameterization is often a hard constraint that may not be correctly specified; this is equivalent to a degenerate prior which violates Cromwell's Rule which, when misspecified, cannot retrieve the true parameter values regardless of sample size.  In what follows, we propose a fully generalizable approach that overcomes these issues.

\subsection{SUBSET priors}
\label{subsec:subset}
\begin{figure}
    \centering
    \begin{subfigure}[b]{0.48\textwidth}
        \includegraphics[width=\textwidth]{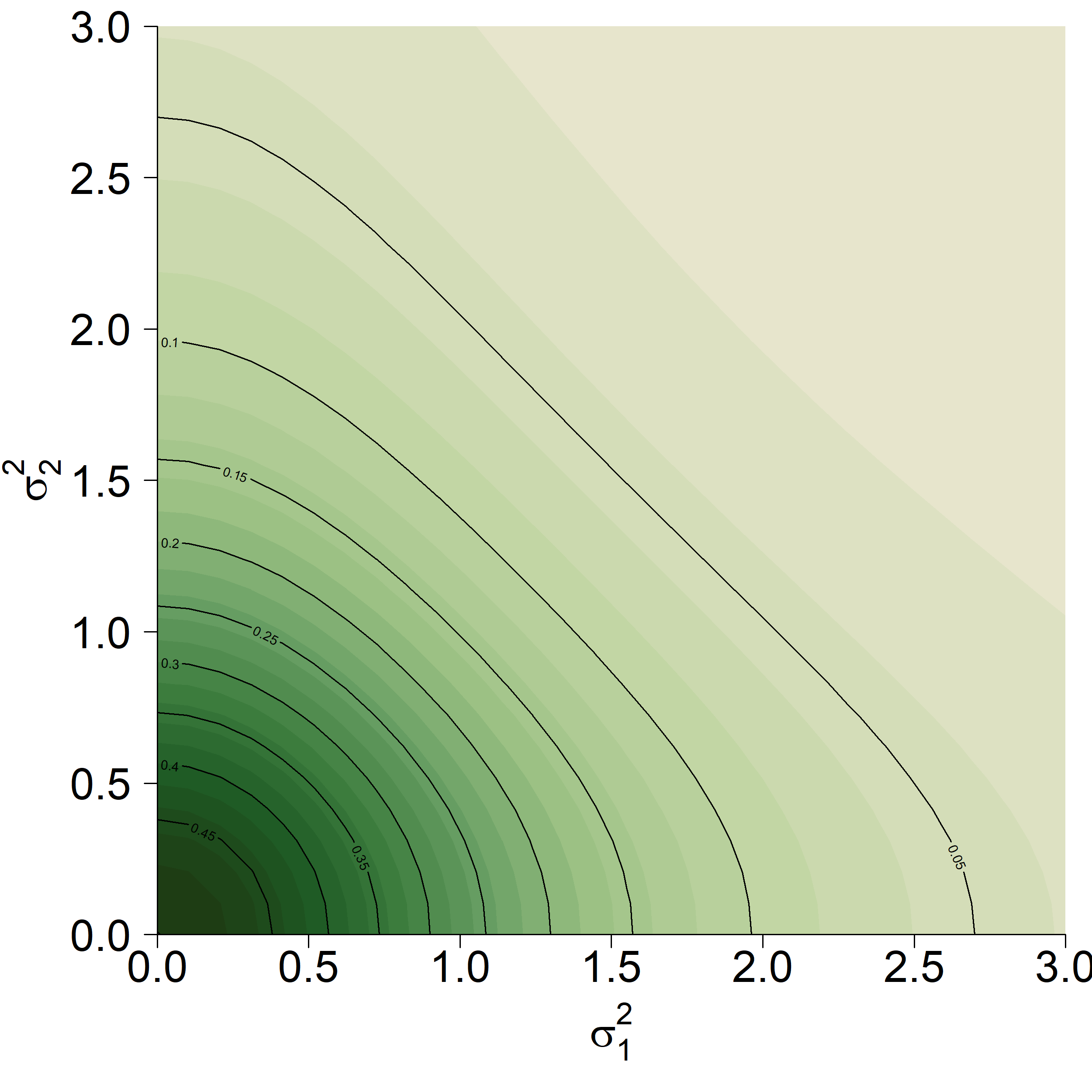}
        \caption{Base prior}
        \label{fig:homoscedastic_example-pi0}
    \end{subfigure}
    \begin{subfigure}[b]{0.48\textwidth}
        \includegraphics[width=\textwidth]{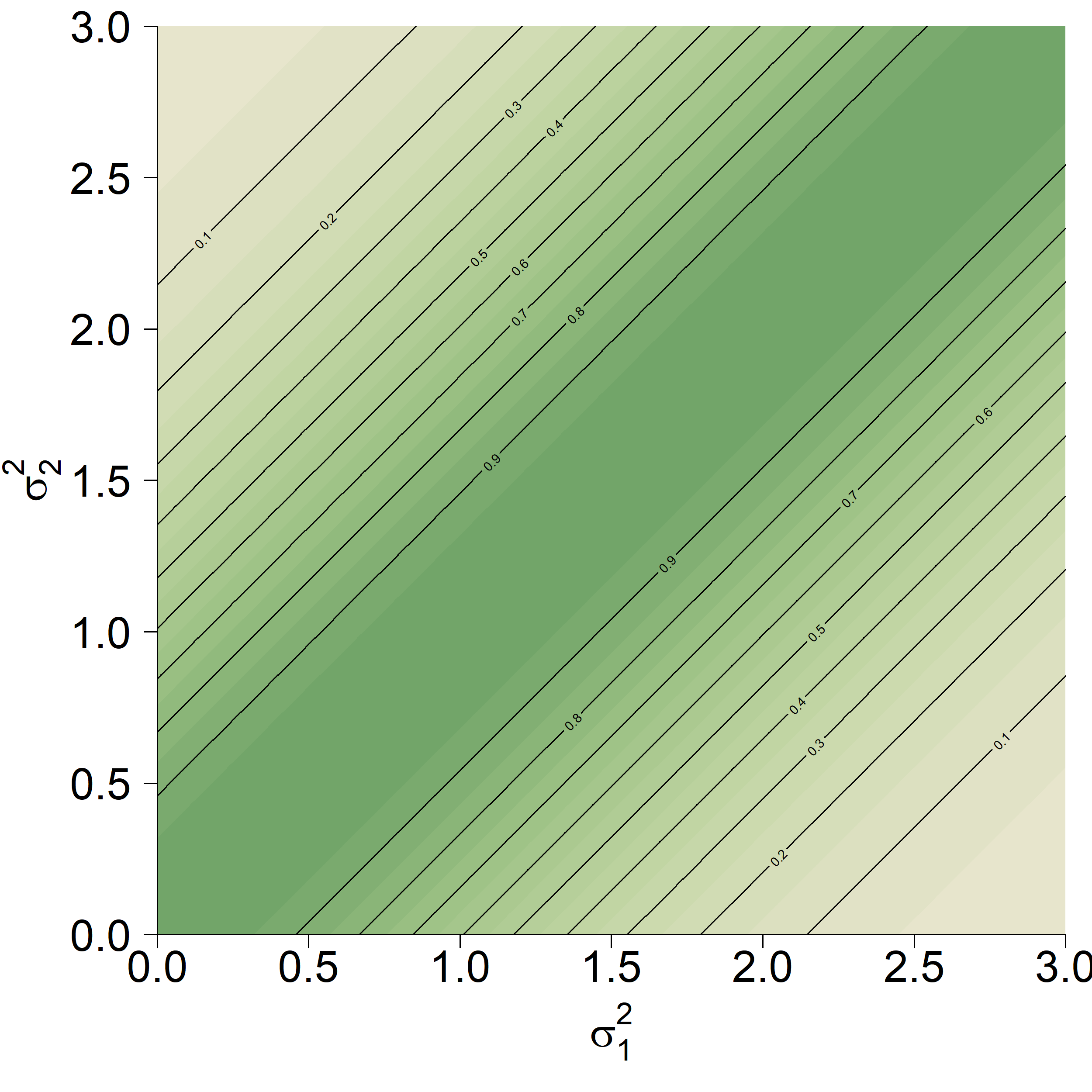}
        \caption{Weighting from exponential tilting term}
        \label{fig:homoscedastic_example-penalty}
    \end{subfigure} \\
    \begin{subfigure}[b]{0.48\textwidth}
        \includegraphics[width=\textwidth]{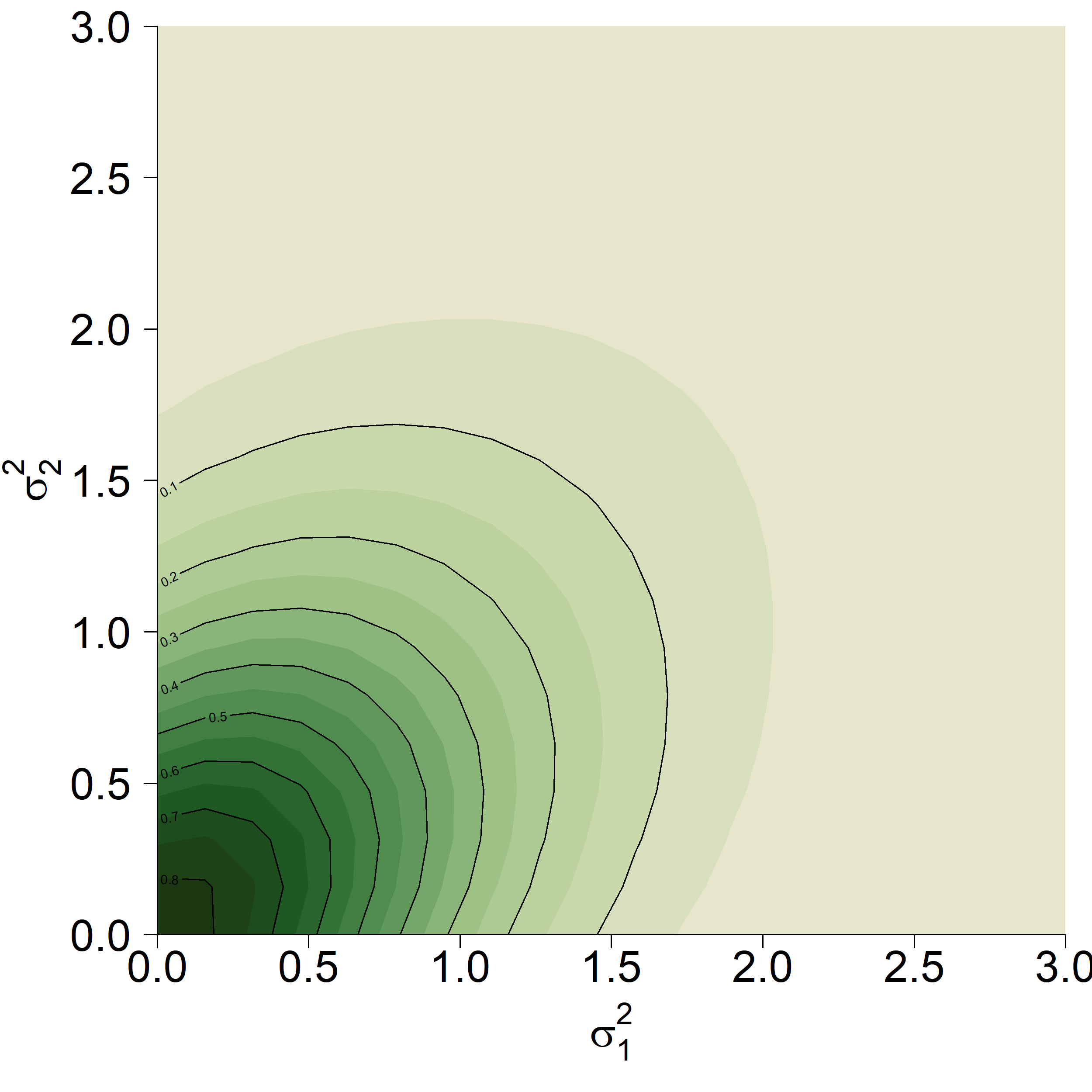}
        \caption{SUBSET prior}
        \label{fig:homoscedastic_example-pi2}
    \end{subfigure}
    \begin{subfigure}[b]{0.48\textwidth}
        \includegraphics[width = \textwidth]{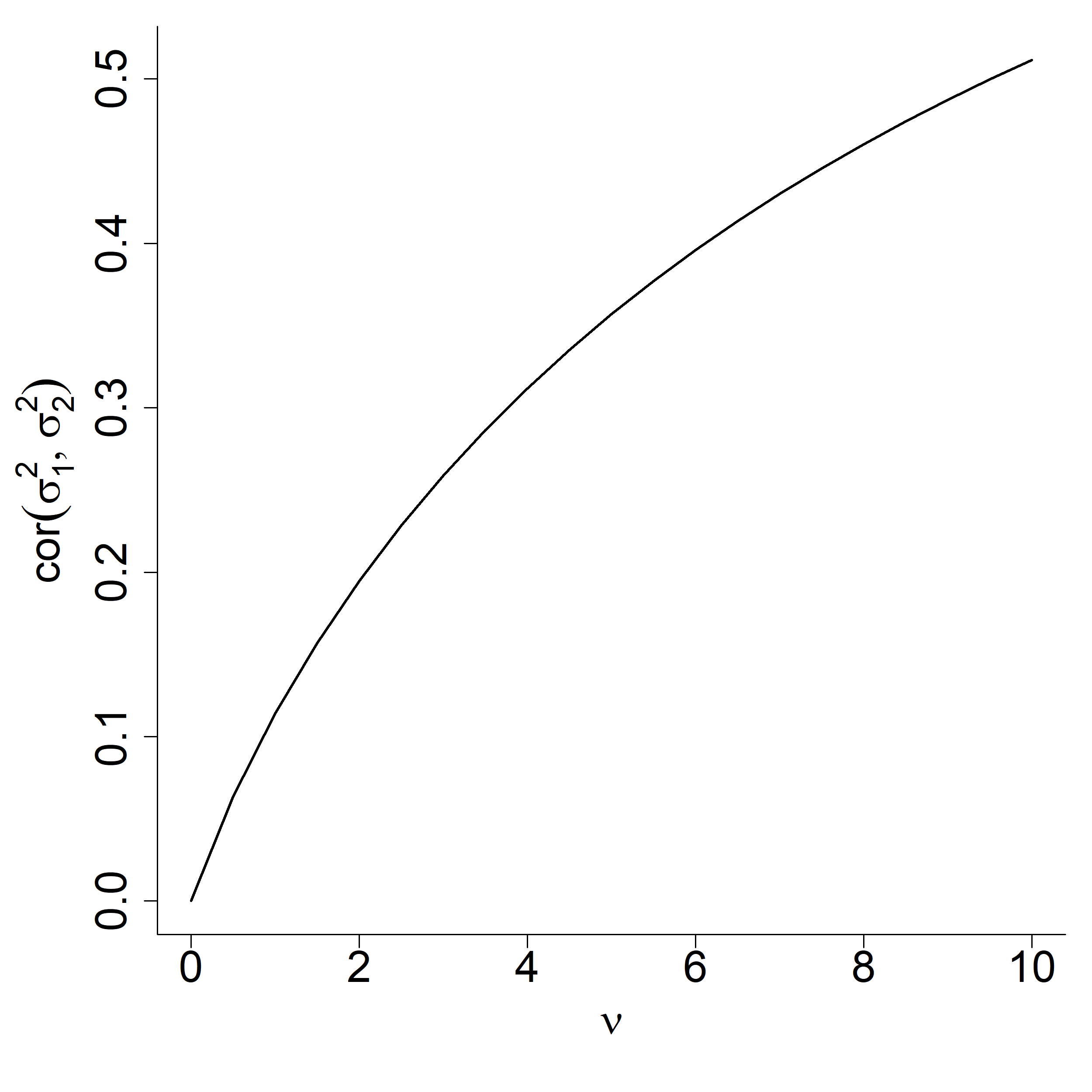}
        \caption{Correlation in $\pi_\nu$ as a function of $\nu$}
        \label{fig:homoscedastic_example-corr}
    \end{subfigure}
    \caption{Illustrative Example \ref{ex:2normal} of applying the SUBSET prior to shrink towards homoscedasticity.  (a) Product of independent half-t distributions with 2 degrees of freedom.  (b) Weighting of the parameter space introduced by the exponential tilting. (c) SUBSET prior using the independent half-t's as the base prior and setting $\nu=2$.  (d) Prior correlation between $\sigma^2_1$ and $\sigma^2_2$ as a function of $\nu$.}
    \label{fig:homoscedastic_example}
\end{figure}

Consider the general case where there is some $L\in\Re^{p\times q}$, such as a design matrix (e.g., $L = (1,1,-1)'$ with $q=1$ in Example \ref{ex:mvnormal} below), that gives rise to a linear subspace $\widetilde{\~L}:=\mbox{span}(L)$, and we believe a priori that $\theta$ lies on or near $\~L:=\widetilde{\~L}\cap \Theta$.  We aim, then, to add this knowledge to our other prior information on $\theta$, i.e., take our base prior $\pi_0$ and adjust it in such a way so as to put smaller prior probability over regions away from $\~L$.  Towards that, we propose using the following exponentially tilted prior $\pi_{\nu}$:
\begin{align}
    \pi_\nu(\theta) & := 
    \frac{1}{Z_{\nu,\phi}}\pi_0(\theta)e^{-\frac{\nu}{2}\theta'(I_p - P(\phi))\theta},& \label{eq:subset} \\
    \nonumber
    \mbox{where } Z_{\nu,\phi} & := \E_{\pi_0}\left(e^{-\frac{\nu}{2}\theta'(I_p - P(\phi))\theta}\right),
\end{align}
and $P(\phi)$ is the projection matrix associated with $\widetilde{\~L}$, which may depend on a user-specified hyperparameter $\phi$. (In \ref{subsec:est_unknown} we will discuss estimating $\phi$, but until that time $\phi$ will be assumed to be a known constant and will not play a role. Thus, we will drop $\phi$ from the notation until that time.)  Eq. (\ref{eq:subset}) shows concretely how we achieve our aim:  $\pi_\nu$ takes the overall shape determined by the base prior $\pi_0$ and through the exponential tilting term penalizes $\pi_0$ in areas which lie on or near the orthogonal subspace of $\~L$ (see Figure \ref{fig:homoscedastic_example-penalty}), implicitly then upweighting areas near $\~L$.  We will henceforth refer to $\pi_\nu$ as the SUBSET (\textbf{SUB}space \textbf{S}hrinkage via \textbf{E}xponential \textbf{T}ilting) prior.  

\begin{proposition}
If $\pi_0$ is a valid probability density function (pdf), the SUBSET prior $\pi_\nu$ is also a valid pdf.
\end{proposition}

The above proposition immediately follows from the fact that $0<e^{-\frac{\nu}{2}\theta'(I_p - P)\theta}\leq 1$ $\forall \theta$. To illustrate the exponentially tilted prior, we provide three examples below. 

\begin{example}
\label{ex:mvnormal}
    Suppose our base prior over $\*\theta = (\theta_1,\theta_2,\theta_3)$ is a multivariate normal distribution centered at zero with spherical covariance matrix, i.e., $\*\theta\sim N(\*0,\tau I_3)$, where $\tau I_3$ is the precision matrix\footnote{In this paper, we will always denote a normal distribution in terms of its precision, rather than variance, so that $N(a,b)$ represents a normal distribution with mean $a$ and variance $b^{-1}$.}. This base prior reflects where we believe $\*\theta$ is centered at ($\*0$), and the degree of uncertainty we have about $\*\theta$ ($\tau$).  Further, suppose that we have reason to believe that $\theta_1$, $\theta_2$, and $\theta_3$ are all roughly equal to each other in magnitude, but that $\theta_3$ is of the opposite sign as $\theta_1$ and $\theta_2$.  Then we have
    \begin{align*}
        \~L & = \mbox{span}\begin{pmatrix} 1 \\ 1 \\ -1 \end{pmatrix}, & &
        P = \frac13 \begin{pmatrix}1 & 1 & -1 \\ 1 & 1 & -1 \\ -1 & -1 & 1  \end{pmatrix}, 
    \end{align*}
    and the SUBSET prior is $\*\theta \sim N\big(\*0, \tau I_3 + \nu(I_3 - P)\big)$, so that $\cor(\theta_1,\theta_2) = \nu/(3\tau + \nu)$ and similarly $\cor(\theta_1,\theta_3) = \cor(\theta_2,\theta_3) = -\nu/(3\tau + \nu)$. We can control the magnitude of the correlation between the variables, then, by increasing or decreasing the exponential tilting parameter $\nu$.
\end{example}

\begin{example}
\label{ex:2normal}
    Suppose we are in a two-sample normal context, i.e., $y_{ij} \overset{ind}{\sim} N(\mu_i,1/\sigma^2_i)$, $i=1,2$, $j=1,\ldots,n_i$.  In this setting it is common to assume homoscedasticity.  This assumption, equivalent to setting $\Pr(\sigma^2_1 \neq \sigma^2_2) = 0$, is highly unlikely to be true in any real context, although it is most often reasonable to assume that there will be near homogeneity, that is, the two variances will be roughly, though not exactly, equal. Instead of making the homogeneity assumption, we can set a base prior on $(\sigma^2_1,\sigma^2_2)$ and shrink this prior towards the linear subspace spanned by $(1,1)'$.  Figure \ref{fig:homoscedastic_example} shows the bivariate base prior set as the product of two independent half-t distributions with 2 degrees of freedom (\ref{fig:homoscedastic_example-pi0}), how that base prior is rescaled due to the exponential tilting term in (\ref{eq:subset}) with $\nu=2$ (\ref{fig:homoscedastic_example-penalty}), and the corresponding SUBSET prior that shrinks towards homoscedasticity (\ref{fig:homoscedastic_example-pi2}).  The hyperparameter $\nu$ controls the degree of shrinkage, which can be seen to dictate the a priori correlation between the two variances in (\ref{fig:homoscedastic_example-corr}).
\end{example}

\begin{figure}
    \centering
    \begin{subfigure}[b]{0.48\textwidth}
        \includegraphics[width = \textwidth]{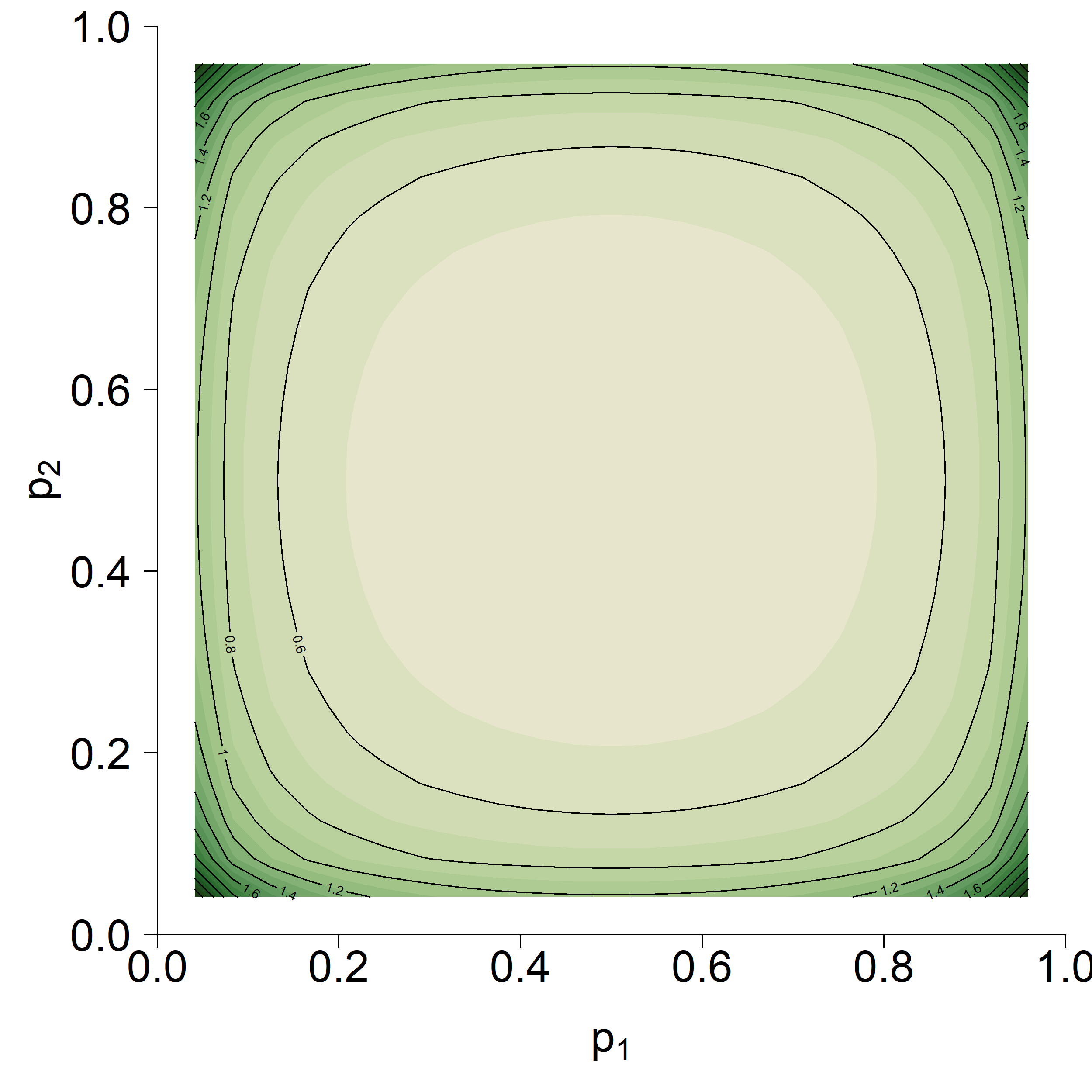}
        \caption{Base prior}
    \end{subfigure}
    \begin{subfigure}[b]{0.48\textwidth}
        \includegraphics[width = \textwidth]{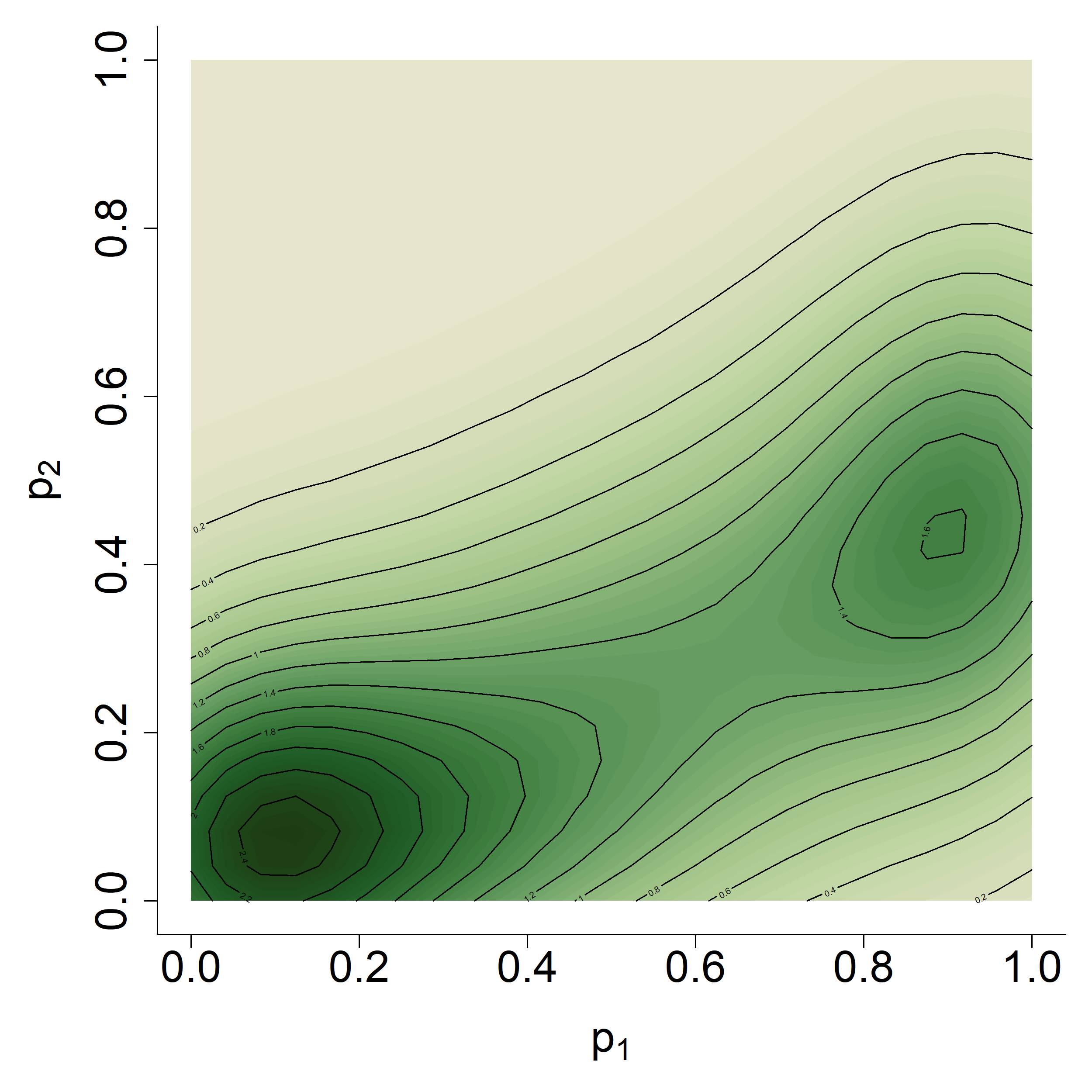}
        \caption{SUBSET prior, $\phi = 2$}
    \end{subfigure}
    \\
    \begin{subfigure}[b]{0.48\textwidth}
        \includegraphics[width = \textwidth]{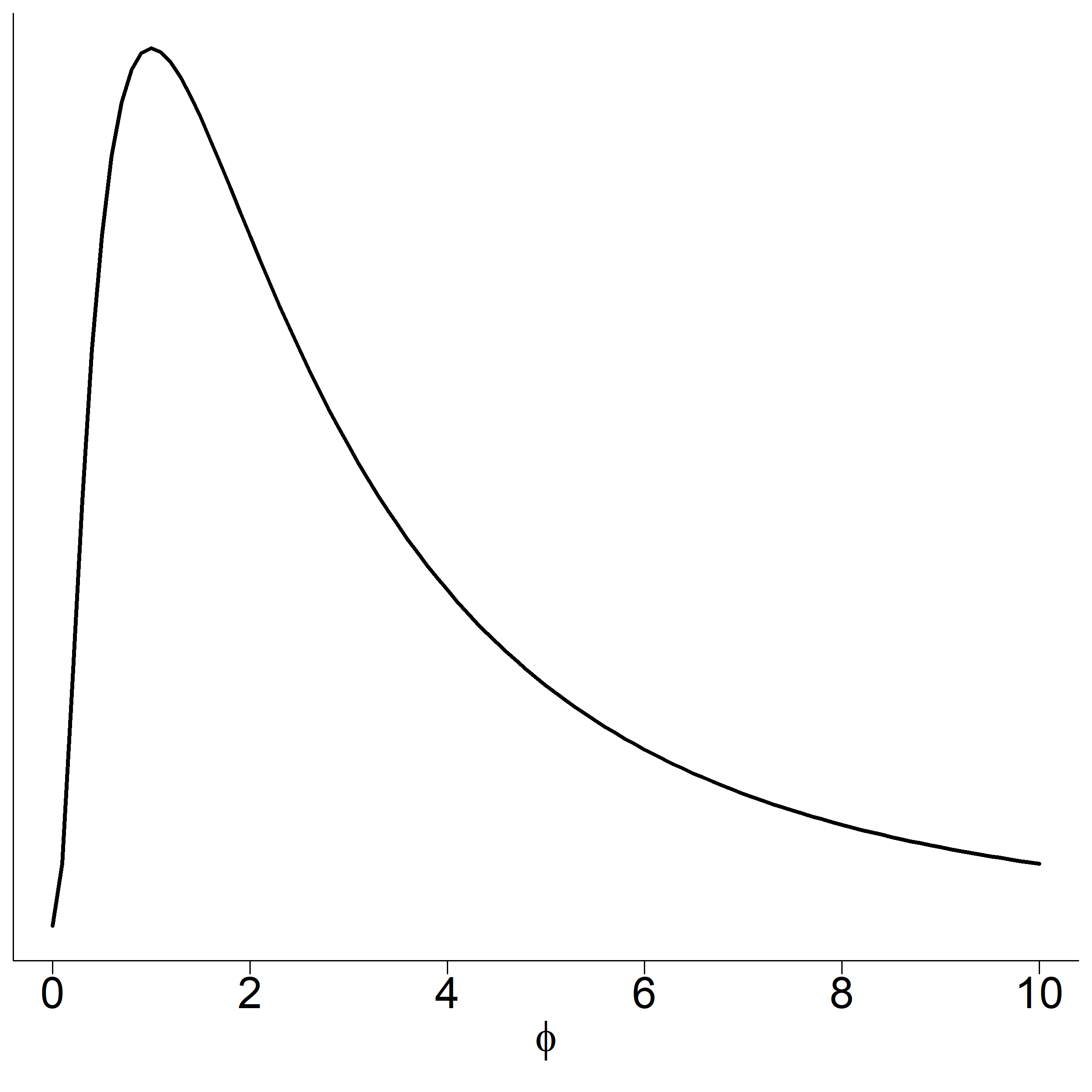}
        \caption{$\phi \sim \ell N(1,1)$}
    \end{subfigure}
    \begin{subfigure}[b]{0.48\textwidth}
        \includegraphics[width = \textwidth]{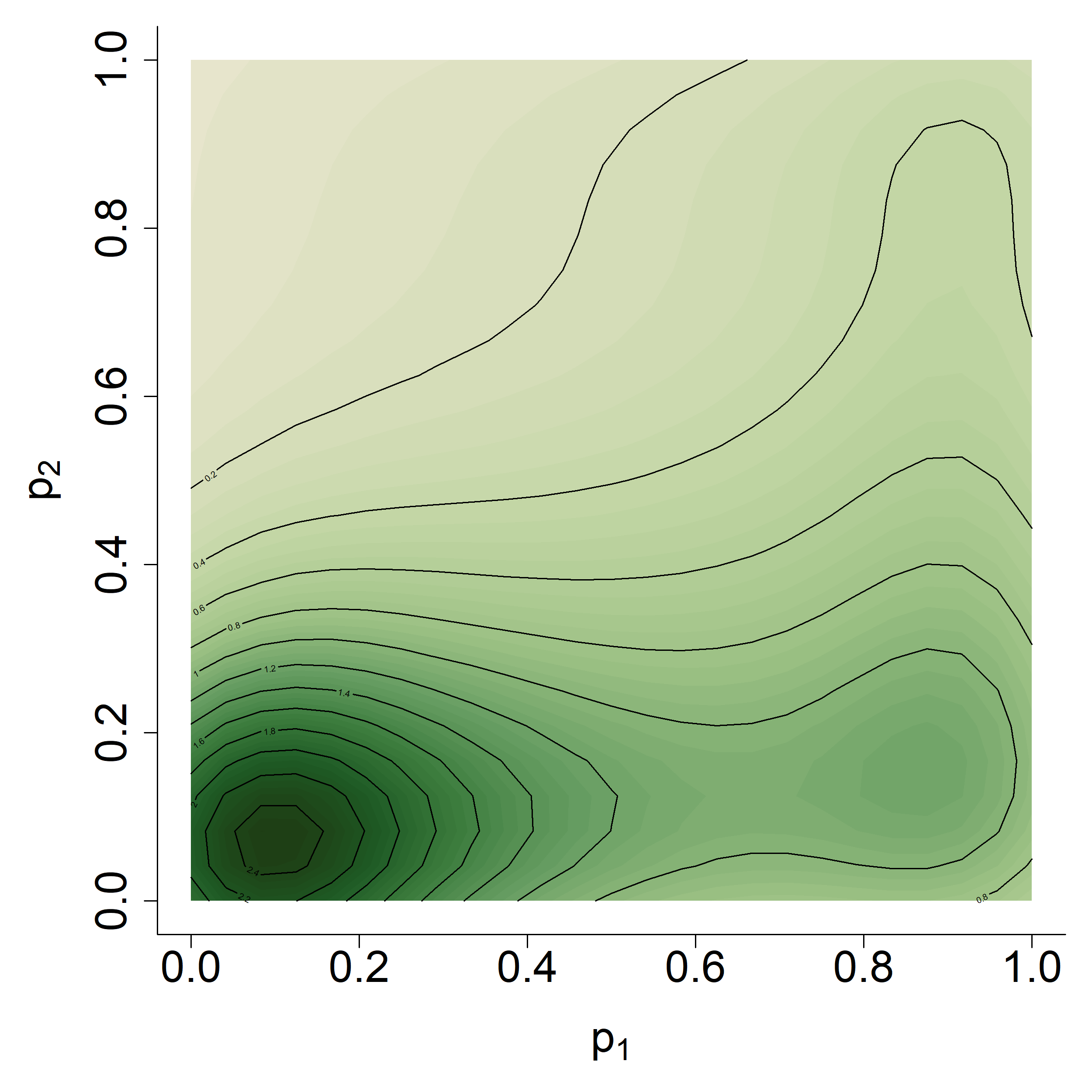}
        \caption{SUBSET prior, unknown $\phi$}
    \end{subfigure}
    \caption{Illustrative Example 3 of applying the SUBSET prior to shrink towards the space where the population 1 response rate is $\phi$ times that of population 2.  The base prior is the product of independent Jeffrey's priors.  In (b) and (d), the tilting parameter $\nu$ was fixed at 50.}
    \label{fig:jeffreys_example}
\end{figure}

\begin{example}
\label{ex:2binomial}
    Suppose we are in the two-sample binomial context, i.e., $y_i\sim Bin(n_i,p_i)$, $i=1,2$.  We use as our base prior a product of independent Jeffreys reference priors, i.e., $Beta(1/2,1/2)$.  However, we believe a priori that it is likely that the response rate $p_1$ is twice that of $p_2$, and hence we wish to push our prior density mass away from regions in $(0,1)^2$ that do not reflect this, i.e., we wish to shrink our prior probability towards $\mbox{span}((2,1)')\cap (0,1)^2$.  Figure \ref{fig:jeffreys_example} shows the Jeffreys priors (a) and the SUBSET prior (c).
\end{example}

\begin{example}
    Consider the heteroscedastic weighted regression model where for $i=1,2,\ldots,N$
    \begin{align*}
        y_i &\overset{ind}{\sim} 
        N(X_{1,i}\beta, 1/\sigma^2_i), 
        & \\
        \sigma^2_i & = 
        X_{2,i}\gamma,
        &
    \end{align*}
    where $X_{1,i}$ and $X_{2,i}$ are covariates used to model the mean and variance structures respectively ($X_{1,i}$ may equal $X_{2,i}$). If we do not wish to make the linear relationship between the regression weights and the covariates $X_{2,i}$ a hard constraint, we may instead consider letting the $\sigma^2_i$ follow, e.g., an inverse gamma distribution, and use a SUBSET prior that shrinks $(\sigma^2_1,\ldots,\sigma^2_N)$ to the column space of $(X_{2,1}',\ldots,X_{2,N}')'$.
\end{example}

The next theorem demonstrates that the SUBSET prior leads to a higher concentration of posterior probability mass near the linear subspace everywhere along that subspace.
\begin{theorem}
\label{th:ball}
    Let $\~L$, $P$, $\pi_0$, and $\pi_{\nu}$ be as defined above. Define $\~B_\epsilon(\~L) := \{\theta:\theta'(I - P)\theta < \epsilon\}$. Let $\pi_0(\theta|y)$ and $\pi_\nu(\theta|y)$ denote the posterior density function of $\theta$ under the base prior and the SUBSET prior respectively, and let $\Pr_0(\theta\in\~A|y)$ and $\Pr_\nu(\theta\in\~A|y)$ denote the posterior probability that $\theta\in\~A$ for some region $\~A\subset\Theta$ under the base and SUBSET priors respectively.    
    
    For any $\~L\neq\emptyset$ and $\epsilon > 0$ such that 
    \[
        0 < \int_{\~B_\epsilon} \pi_j(\theta|y)d\theta < 1, \hspace{2pc} j\in\{0,\nu\},
    \]
    we have
    \begin{equation*}
        \mathrm{Pr}_{\nu}(\theta \in \~B_\epsilon|y) > \mathrm{Pr}_0(\theta \in \~B_\epsilon|y).
    \end{equation*}
\end{theorem}
The proof can be found in the Supplementary Material \citep{sewell2023supplementLSS}.

By influencing the posterior through the prior, all posterior inference, not just point estimation, is affected by the prior belief that $\theta$ lives on or near the linear subspace.  Point estimation, too, is affected by using a SUBSET prior, but through the appropriate mechanism of affecting the Bayes risk through the posterior.

\vspace{1pc}
\noindent \textbf{Remark 1.} \cite{Lee1994shrinking} proposed a ``subspace ridge'' estimator, $\hat\beta_{SRDG}$, in a linear regression setting ($y|X\beta,\sigma^2 \sim N(X\beta,I_n / \sigma^2)$) which shrinks the estimates of $\beta$ towards a linear subspace with projection matrix $P$.  If the commonly implemented improper prior $\pi(\beta,\sigma^2)\propto 1/\sigma^2$ is used as a base prior and $\nu$ is set to $k/\sigma^2$ for some $k>0$, then the posterior mean obtained from the SUBSET prior would be equivalent to the subspace ridge estimator, namely
\[
    \E_{\pi_\nu}(\beta|y) = 
    \big(X'X + k (I-P)\big)^{-1}X'y = \hat\beta_{SRDG}.
\]
Note that the focus of \cite{Lee1994shrinking} was on point estimation, rather than obtaining a posterior distribution which incorporated the prior knowledge that the regression coefficients ought to lie on or near the subspace.

\vspace{1pc}
\noindent \textbf{Remark 2.} It is easy to accommodate shrinkage of subsets of parameters towards different subspaces, as well as have some parameters not shrunk at all.  That is, suppose we have a partition of size $(R+1)$ of a $p$-dimensional set of parameters $\theta = (\theta_0,\theta_1,\theta_2,\ldots,\theta_R$), where $\theta_0$ of dimension $p_0$ is not being shrunk towards a linear subspace, and for $r=1,\ldots,R$, $\theta_r$ of dimension $p_r$ is being shrunk towards the linear subspace with $p_r\times p_r$ projection matrix $P_r$. Then we can set $P$ in Equation (\ref{eq:subset}) to be the block diagonal matrix with diagonal blocks equal to $I_{p_0}$, $P_1$, \ldots, $P_R$.

\section{Estimation}
\label{sec:estimation}
\subsection{Fixed $\~L$}
\label{subsec:est_fixed}
\green{Posterior inference can, of course, be accomplished in all of the usual ways.  However, as we will describe shortly, when shrinking the posterior mass towards a linear subspace it is beneficial to evaluate the effect of various levels of shrinkage, i.e., values of $\nu$, as well as to compare the results to those obtained from no shrinkage.  With this in mind, it becomes clear that an efficient estimation method is needed.  That is, in many- if not most- real data analyses, posterior sampling of $\theta|y$ is computationally expensive, and if MCMC is used convergence diagnostics must be obtained; performing this repeatedly for many values of $\nu$ quickly becomes untenable in most situations.  The following importance sampler overcomes this issue by only requiring posterior samples of $\theta|y$ under the base prior to be obtained once, followed by very fast importance sampling for each value of $\nu$ whose computational cost does not, for example, scale with the sample size nor with the length of a MCMC burn-in period.}

\subsubsection{Importance sampler}
\label{subsubsec:est_fixed_is}
Suppose we have obtained posterior samples $\{\theta_{\green{y,}k}\}_{k=1}^{K_{\green{y}}}$ under the base prior, i.e., from 
\begin{equation}
\pi_0(\theta|y) \propto \pi(y|\theta)\pi_0(\theta).
\end{equation}
\green{as well as samples $\{\theta_{0,k}\}_{k=1}^{K_0}$ taken directly from the base prior $\pi_0$.} We propose using \green{$\pi_0(\cdot|y)$} as the importance distribution to obtain samples from the posterior under the SUBSET prior
\begin{equation}
\pi_\nu(\theta|y) \propto \pi(y|\theta)\pi_0(\theta)e^{-\frac{\nu}{2}\theta'(I-P)\theta}.
\end{equation}

Under this importance distribution, the \green{unnormalized} importance weights $\green{w_{y,k}(\nu)}$ corresponding to \green{$\theta_{y,k}$}, \green{are simply}
\begin{align}
    \green{w_{y,k}(\nu)} 
    & \propto 
    \exp\left\{
        -\green{\frac{\nu}{2}}\theta_{\green{y,}k}'(I-P)\theta_{\green{y,}k}
    \right\}.
    \label{eq:is_fixed_weights}
\end{align}

\green{In choosing $\nu$, one has (at least) three choices.  First, $\nu$ can be selected manually a priori based on obtaining a prior distribution over $\theta$ matching one's prior beliefs.  

Second, one can do a sensitivity analysis a posteriori, evaluating the effects of various values of $\nu$.  This is very fast to compute since (1) $w_{y,k}(\nu)$ does not depend on sample size, and (2) $w_{y,k}(\nu) = [w_{y,k}(1)]^\nu$, and hence (\ref{eq:is_fixed_weights}) need only be computed once for $\nu=1$, and then these unnormalized weights can be exponentiated to obtain the weights for other values of $\nu$.  If one were to take this approach, it is important to ensure that the effective sample size does not decrease below an acceptable threshold, for as $\nu$ increases, the posterior under the SUBSET prior will shift farther away from $\pi_0(\cdot|y)$ leading to weight degeneracy. 

Third, one can use Bayes factors to determine the value of $\nu$. The Bayes factor for the SUBSET prior vs. base prior can be written as 
\begin{equation}
    \frac{\pi_\nu(y)}{\pi_0(y)} = \frac{\E_{\pi_0}\left(e^{-\frac{\nu}{2}\theta'(I-P)\theta}\Big|y\right)}{\E_{\pi_0}\left(e^{-\frac{\nu}{2}\theta'(I-P)\theta}\right)}
    \label{eq:bf}
\end{equation}
\citep[see Supplementary Material for derivation;][]{sewell2023supplementLSS}.  This can then be approximated via 
\begin{equation}
    \frac{\pi_\nu(y)}{\pi_0(y)} \approx \frac{
  \frac{1}{K_y}\sum_{k=1}^{K_y} w_{y,k}(\nu)
  }{
  \frac{1}{K_0}\sum_{k=1}^{K_0} w_{0,k}(\nu)
  },
  \label{eq:bf_approx}
\end{equation}
where $w_{0,k}(\nu):= \exp\{-\frac\nu2\theta_{0,k}'(I-P)\theta_{0,k}\}.$
Again, since $w_{0,k}(\nu) = [w_{0,k}(1)]^\nu$ and $w_{y,k}(\nu) = [w_{y,k}(1)]^\nu$, it is relatively fast to numerically maximize (\ref{eq:bf_approx}) as a function of $\nu$. This importance sampling algorithm using Bayes factor to select $\nu$ is provided in Algorithm \ref{alg:is_fixed}.
}

\begin{algorithm}
\SetKwFunction{helper}{helper}
\green{
\KwIn{Posterior samples $\{\theta_{y,k}\}_{k=1}^{K_{y}}$ under the base prior; Prior samples $\{\theta_{0,k}\}_{k=1}^{K_0}$; Projection matrix $P$.}

\For{$k=1,\ldots,K_0$}{
Compute $w_{0,k}(1) = \exp\{-\frac12\theta_{0,k}'(I-P)\theta_{0,k}\}$ 
}
\For{$k=1,\ldots,K_y$}{
Compute $w_{y,k}(1) = \exp\{-\frac12\theta_{y,k}'(I-P)\theta_{y,k}\}$ 
}

 \SetKwFunction{FMain}{helper}
    \SetKwProg{Fn}{Function}{:}{}
    \Fn{\FMain{$\nu$}}{
    \textbf{return} $
        \frac{
        \frac{1}{K_y}\sum_{k=1}^{K_y} [w_{y,k}(1)]^\nu
        }{
        \frac{1}{K_0}\sum_{k=1}^{K_0} [w_{0,k}(1)]^\nu
        }
    $
    }

\vspace{0.5pc}
Use univariate optimizer to find $\nu^* := \underset{\nu}{\mbox{argmax}}$ \helper$(\nu)$

Compute $w_{y,k}(\nu^*) := [w_{y,k}(1)]^{\nu^*}$

\KwRet{$\{\theta_{y,k},w_{y,k}(\nu^*)\}_{k=1}^{K_y}$}
}
\caption{\green{Importance sampler for posterior sampling under the SUBSET prior, returning the draws from the importance distribution and their importance weights, where the shrinkage parameter $\nu$ is selected by largest Bayes factor.}}
\label{alg:is_fixed}
\end{algorithm}

\subsubsection{Large sample approximation}
\label{subsubsec:est_fixed_asymp}
For large sample sizes (where posterior sampling can become tedious), many important and historical works have proved the asymptotic normality of the posterior distribution \citep{bernstein1917theory}. 
While certain settings have required further development of these theorems \citep[e.g.,][develops such results for semi- and non-parametric posteriors]{shen2002asymptotic}, fairly general conditions for where this holds are given in \cite{chen1985asymptotic} and will be our focus moving forward.

\newpage
If the conditions outlined by \citeauthor{chen1985asymptotic} hold, then under $\pi_0$
\[
\theta|y,\pi_0 \overset{\cdot}{\sim} N(m_n,\Omega_n), 
\]
where $m_n$ is the posterior mode and $\Omega_n$ is the Hessian of the negative log posterior at $m_n$.  In this setting, we can approximate the exponentially tilted posterior in the following way:
\begin{align} \nonumber
    \pi_{\nu}(\theta|y) & \propto 
    \pi(y|\theta)\pi_0(\theta)e^{-\frac\nu2\theta'(I_p - P)\theta} &\\ \nonumber
    & \overset{\cdot}{\propto}
    \exp\left\{
    -\frac12\Big[
        \theta'\Big(\Omega_n + \nu(I_p-P)\Big)\theta - 2\theta'\Omega_n m_n
    \Big]
    \right\}, & \\
    \Rightarrow \theta|y,\pi_{\nu} & \overset{\cdot}{\sim} N(\tilde m_n, \widetilde\Omega_n),& \label{eq:normal_fixed_phi} \\ \nonumber
    \widetilde\Omega_n & := \Omega_n + \nu(I_p - P),&\\ \nonumber
    \tilde m_n & := \widetilde\Omega_n^{-1}\Omega_n m_n.&
\end{align}
This implies that once the posterior mode and Hessian under the base prior are computed (either analytically, numerically, or via posterior sampling), there is negligible computational cost required to obtain an approximate posterior under exponential tilting with tilting parameter $\nu$.  

\green{
The selection of $\nu$ using Bayes factors can be done in a similar fashion to Algorithm \ref{alg:is_fixed}.  The only change necessary is to replace the numerator in the \texttt{helper} function to be the expectation evaluated analytically:
\begin{align*}
\E_{\pi_0}\left(
e^{-\frac{\nu}{2}\theta'(I-P)\theta}\big|y
\right) 
& =|\Omega_n|^{\frac 12}|\widetilde \Omega_n|^{-\frac 12}
\exp\left\{
-\frac12\left(
m_n'\Omega_nm_n - \tilde m_n'\widetilde \Omega_n \tilde m_n
\right)
\right\}.
\end{align*}
}

\subsection{Estimating $\~L$}
\label{subsec:est_unknown}
The subspace towards which we should shrink our posterior is not always precisely defined.  Continuing Example \ref{ex:2binomial}, we may be confident that $p_1$ is larger than $p_2$, and while we may be confident that $p_1$ is somewhere in the neighborhood of 2 times larger than $p_2$, we might be hard pressed to specify that factor exactly.  

Consider a more general setting, where the subspace is the span of some column vectors which in turn are a function of unknown parameters $\phi$ taking values in $\Phi$, and let $P(\phi)$ denote the corresponding projection matrix.  In the preceding example we were considering $\~L= \mbox{span}((\phi,1)')\cap (0,1)^2$ for $\phi=2$, but we may relax this so that $\phi\in(0,\infty)$. Figure \ref{fig:jeffreys_example} shows a log-normal prior over $\phi$ (c) and the resulting marginal prior over the two population response rates (d).

In the case of unknown $\phi$, a 2-block Metropolis-Hastings-within-Gibbs sampler can be effectively employed to obtain posterior samples in a computationally efficient manner, where we alternate between sampling $\theta$ and sampling $\phi$.  This is due to the fact that for fixed $\theta$, if we draw $\phi^*$ from its prior independently from our Markov chain's current value $\phi^{curr}$, it can be shown that a Metropolis-Hastings sampling step accepts $\phi^*$ with probability
\begin{equation}
    \min\left\{
        1,
        \exp\left\{-\frac{\nu}{2}{\theta^{(k)}}'(P(\phi^{curr}) - P(\phi^*))\theta^{(k)}\right\}\cdot\frac{Z_{\nu,\phi^{curr}}}{Z_{\nu,\phi^*}}
    \right\}.
    \label{eq:phi_accProb}
\end{equation}

\subsubsection{Importance sampler (unknown $\phi$)}
\label{subsubsec:est_phi_is}
Let $\phi$ take a finite number of values with corresponding prior probability mass function $\pi_\phi$. In the case where $\phi$ is better considered continuous with probability density function $\tilde \pi_\phi$, given that $\phi$ will not be of primary importance, we assume that we may sufficiently approximate this by taking a sequence of quantiles $\{\phi_q\}_{q=1}^Q$ and taking $\pi_\phi(\phi_q) \propto \tilde \pi_\phi(\phi_q)$.

Unlike the case with fixed $\phi$, now we must worry about the SUBSET normalizing constant $Z_{\nu,\phi}$, which typically will not have a closed form solution.  However, since $Z_{\nu,\phi}$ is an expectation with respect to the base prior on $\theta$, we may obtain an arbitrarily accurate Monte Carlo estimate $\widehat Z_{\nu,\phi}$ by taking samples from  $\pi_0(\theta)$, which will typically be easy to accomplish (as was done in Algorithm \ref{alg:is_fixed}).  Hence in our 2-block Gibbs sampler, for a given $\theta$ we can draw $\phi^*$ from $\pi_\phi$ and accept it with probability given in Eq. (\ref{eq:phi_accProb}), substituting $Z_{\nu,\phi}$ with its Monte Carlo estimate $\widehat Z_{\nu,\phi}$.

For a fixed $\phi\in\{\phi_q\}_{q=1}^Q$, we can implement the importance sampler described in Section \ref{subsec:est_fixed} to approximate the full conditional of $\theta|y,\phi$ by the atomized approximation
\begin{equation}
    \hat\pi(\theta|y,\phi=\phi_q) = \sum_{k=1}^{K_{\green{y}}} \tilde w_{(\phi_q)\green{y,}k}\delta_{\green{\theta_{y,k}}}(\theta),
\end{equation}
where \green{$\{\theta_{y,k},w_{(\phi_q)\green{y,}k}\}_{k=1}^K$} are the samples and importance weights obtained through fixing $\phi=\phi_q$ and \green{computing the weights via Eq. \ref{eq:is_fixed_weights}}, and $\tilde w_{(\phi_q)\green{y,}k}$ is the normalized importance weight for the $k^{th}$ sample.  The importance sampler is extremely fast, and draws from the importance distribution along with their weights can be obtained for each of the finite values of $\phi$ before running the Gibbs sampler. The resulting 2-block Gibbs sampler is described in Algorithm \ref{alg:is_gibbs}.

\begin{algorithm}
    \KwIn{Posterior samples $\{\theta_{\green{y},k}\}_{k=1}^{K_{\green{y}}}$ under the base prior; Prior samples $\{\theta_{0,k}\}_{k=1}^{K_0}$; Projection matrix function $P(\cdot)$; Initial values $\theta_{\nu,0}$ and $\phi_{\nu,0}$; Shrinkage weight $\nu$; Prior $\pi_\phi$ over the set $\{\phi_q\}_{q=1}^Q$; \green{Desired number of posterior samples $K_\nu$.}}

    \vspace{1pc}
    \tcc{Precompute key quantites for Gibbs sampler}
    
    \For{$q=1,\ldots,Q$}{
        Compute $\widehat Z_{\nu,\phi_q} = \frac{1}{K_{\green{0}}}\sum_{k=1}^{K_{\green{0}}} e^{-\frac\nu2\theta_{0,k}'(I-P(\phi_q))\theta_{0,k}}$

        \green{Compute $w_{(\phi_q)y,k} = e^{-\frac\nu2\theta_{y,k}'(I-P(\phi_q))\theta_{y,k}}$ for $k=1,\ldots,K_y$}
    }

    \vspace{1pc}
    Compute $P^{curr} \leftarrow P(\phi_{\nu,0})$
    
    \vspace{1pc}
    \tcc{Perform 2-block Gibbs sampler}
    \For{$k=1,\ldots,K_{\green{\nu}}$}{
        \tcc{Draw a new $\theta|y,\phi$}
        Set $\theta_{\nu,k} = \green{\theta_{y,k'}}$ with probability proportional to $w_{(\phi_{\nu,k-1})\green{y,}k'}$ 

        \vspace{1pc}
        \tcc{Draw a new $\phi|y,\theta$}
        Draw $\phi^*$ from \green{the prior over $\phi$ ($\pi_\phi$)}
        
        Compute $P^* \leftarrow P(\phi^*)$
        
        Draw $u\sim Unif(0,1)$
        
        \eIf{$u < \exp\left\{-\frac{\nu}{2}\theta_{\nu,k}'(P^{curr} - P^*)\theta_{\nu,k}\right\}\cdot\frac{\widehat Z_{\nu,\phi_{\nu,k-1}}}{\widehat Z_{\nu,\phi^*}}$}{
            
            $P^{curr} \leftarrow P^*$
            
            $\phi_{\nu,k} \leftarrow \phi^*$;\\
    }{
        $\phi_{\nu,k} \leftarrow \phi_{\nu,k-1}$
    }
    }

    \KwRet{$\{\theta_{\nu,k}, \phi_{\nu,k}\}_{k=1}^{K_{\green{\nu}}}$}
    
    \caption{Two-block MH-within-Gibbs sampler for \green{unknown $\theta$ and} unknown $\phi$ (i.e., unknown linear subspace), relying on importance sampling.}
    \label{alg:is_gibbs}
\end{algorithm}

\subsubsection{Large sample approximation (unknown $\phi$)}
\label{subsubsec:est_phi_asymp}

Suppose that the posterior arising from the base prior can again be well approximated with a normal distribution.  Given a fixed value of $\phi$, the full conditional distribution of $\theta$ is the normal
distribution given in Eq. (\ref{eq:normal_fixed_phi}). As in Algorithm \ref{alg:is_gibbs}, a simple Metropolis-Hastings step can be taken to update $\phi$.  

Since we no longer need to perform importance sampling for each value of $\phi$, we no longer constrain $\phi$ to take values on a discrete set. Yet the need to compute Monte Carlo estimates of the normalizing constant $Z_{\nu,\phi}$ can still slow the sampling algorithm to a debilitating level. However, as the next proposition shows, under arguably most scenarios we have smoothness in $Z_{\nu,\phi}$ which can aid computation significantly.

\begin{theorem}
\label{prop:continuity}
Suppose $L=L(\phi)$ is of full rank and differentiable in $\phi$ for all $\phi\in\Phi$, and $P(\phi):=L\big(L'L\big)^{-1}L'$.  Then for any $\nu>0$, $Z_{\nu,\phi}$ is continuous in $\phi$ over $\Phi$.
\end{theorem}
The proof is in the Supplementary Material \citep{sewell2023supplementLSS}.

\noindent We propose, therefore, prior to performing the 2-block Gibbs sampler to perform a two-stage estimation scheme for $Z_{\nu,\phi}$, using a spline regression- or if $\phi$ is multivariate use tensor product splines or thin-plate splines- for a sequence of values of $\phi$ predicting $\hat Z_{\nu,\phi}$, and then in the Gibbs sampler using estimated values of $Z$ from this spline fit.  This approach is given in Algorithm \ref{alg:gibbs_spline}.

\begin{algorithm}
\KwIn{Posterior mode $m_n$ and precision matrix $\Omega_n$ using the base prior; Prior samples $\{\theta_{0,k}\}_{k=1}^{K_{\green{0}}}$; Projection matrix function $P(\cdot)$; Initial values $\theta_{\nu,0}$ and $\phi_{\nu,0}$; Shrinkage weight $\nu$; Prior $\pi_\phi$; Spline function $f:\Phi\mapsto \Re$; Sequence $\{\phi_s\}_{s=1}^S$; Number of posterior draws $K_{\green{\nu}}$.}

    \vspace{1pc}
    \tcc{Precompute key quantites for Gibbs sampler}
    Compute $P^{curr} \leftarrow P(\phi_{\nu,0})$

    \For{$s = 1,\ldots,S$}{
        Compute $\widehat Z_{\nu,\phi_s} = \frac1{K_{\green{0}}}\sum_{k=1}^{K_{\green{0}}} e^{-\frac\nu2 {\theta_{0,k}}'(I-P(\phi_s))\theta_{0,k}}$
    }
    
    Fit spline model $\skew{3}\widehat {\widehat Z}(\cdot)$ from regressing $\widehat Z_{\nu,\phi_s}$ on $f(\phi_s)$
    
    \vspace{1pc}
    \tcc{Perform 2-block Gibbs sampler}
    \For{$k = 1,\ldots,K_{\green{\nu}}$}{
        \tcc{Draw a new $\theta|y,\phi$}
        Draw $\theta_{\nu,k}$ from $N(\tilde m_n,\widetilde\Omega_n)$ as defined in Eq. (\ref{eq:normal_fixed_phi}) using $P=P(\phi_{\nu,k-1})$
        
        \vspace{1pc}
        \tcc{Draw a new $\phi|y,\theta$}
        Draw $\phi^*$ from \green{the prior over $\phi$ ($\pi_\phi$)}
        
        Compute $P^* \leftarrow P(\phi^*)$
        
        Draw $u\sim Unif(0,1)$
        
        \eIf{$u < \exp\left\{-\frac{\nu}{2}\theta_{\nu,k}'(P^{curr} - P^*)\theta_{\nu,k}\right\}\cdot\frac{\skew{3}\widehat {\widehat Z}\left(\phi_{\nu,k-1}\right)}{\skew{3}\widehat {\widehat Z}\left(\phi^*\right)}$}{
            $P^{curr} \leftarrow P^*$;\\
            $\phi_{\nu,k} \leftarrow \phi^*$
        }{
            $\phi_{\nu,k} \leftarrow \phi_{\nu,k-1}$
        }
    }

    \KwRet{$\{\theta_{\nu,k}, \phi_{\nu,k}\}_{k=1}^{K_{\green{\nu}}}$}
    
    \caption{Two-block MH-within-Gibbs sampler for \green{unknown $\theta$ and} unknown $\phi$ (i.e., unknown linear subspace), relying on a large sample approximation of the posterior. Note that if $\Phi$ is finite, $\{\phi_s\}_{s=1}^S$ should equal $\Phi$, and $\skew{3}\widehat {\widehat Z}(\phi) = \widehat Z_{\nu,\phi}$ rather than predictions from the spline fit.}
    \label{alg:gibbs_spline}
\end{algorithm}

\section{Simulation studies}
\label{sec:simstudy}
To evaluate the strengths and weaknesses of the usage of SUBSET priors, we ran two simulation studies.  The results shown here correspond to the importance samplers; using the large sample approximations yielded similar results, which can be found in the Supplementary Material \citep{sewell2023supplementLSS}.

\subsection{1-way ANOVA}
\label{subsec:simstudy_anova}
Our first simulation study used a 1-way ANOVA setting, evaluating the SUBSET prior on the estimation and inference of the group variances.  The data we simulated used 6 groups, each with 20 observations, for a total sample size of 120, i.e., 
\[
    y_{gi} \overset{iid}{\sim} N(\mu_g,1/\sigma^2_g),\hspace{2pc} g=1,\ldots,6,\hspace{1pc} i=1,\ldots,20.
\]
The group means were set to be $(1,2,\ldots,6)$, and the group variances were set according to three scenarios:
\begin{itemize}
    \item \textit{Homoscedasticity.} Each group had a residual variance of 2.
    \item \textit{Mild heteroscedasticity.} The group variances were $(1, 1.6, 2.2, 2.8, 3.4, 4)$.
    \item \textit{Strong heteroscedasticity.} The group variances were $(1,3,5,7,9,11)$.
\end{itemize}
Variances, even when not of interest directly, are important for other quantities of interest, e.g., prediction intervals and Exceedance in Pairs Rate\footnote{Roughly, EPR is a measure of how well separated the groups are.} \citep{rosner2021bayesian}.

We fit four models to each data set. The first assumed homoscedasticity, using a Normal-Inverse gamma prior on the means and (common) variance, i.e.,
\begin{align*}
    \mu_g |\sigma^2 
    & \overset{iid}{\sim} 
    N(0,a/\sigma^2), 
    &\\
    \sigma^2 
    & \sim
    \Gamma^{-1}(b/2,c/2),
\end{align*}
We set $a=1$, $b=1$, and $c=2$.

The second model assumed heteroscedasticity, with a prior structure similar to that given above:
\begin{align*}
    \mu_g |\sigma^2_g 
    & \overset{ind}{\sim}
    N(0,a/\sigma^2_g), 
    &\\
    \sigma^2_g
    & \overset{iid}{\sim}
    \Gamma^{-1}(b/2,c/2).
\end{align*}

The third model used the heteroscedastic prior given above as the base prior for a SUBSET prior with \green{$\nu$ selected via Bayes factor according to Algorithm \ref{alg:is_fixed}.} We shrunk the group variances towards the subspace spanned by $(1,\ldots,1)'$, i.e., towards homoscedasticity.

Each model used 50000 posterior draws, and data from each of the three scenarios were generated and analyzed 2000 times.

Table \ref{tbl:simstudy_anova} provides the results for the estimation of $\{\sigma^2_g\}_{g=1}^6$ in terms of \green{95\%} credible interval (CI) widths, \green{95\%} CI coverage rates, and MSE.  In the case of homoscedastic data, fitting the homoscedastic model unsurprisingly yields the lowest CI widths and lowest MSE.  All approaches yield coverage rates near the nominal level.  \green{Importantly, the SUBSET prior improves the performance of the heteroscedastic model in terms of CI width, CI coverage, and MSE. For mildly heteroscedastic data, the homoscedastic model cannot appropriately model the uncertainty due to the hard constraint of homoscedasticity, leading to very low coverage and high MSE. However, compared to the heteroscedastic model, the SUBSET prior yields a 26\% reduction in MSE and a 29\% reduction in the average CI width, although at a cost of a 0.07 reduction in coverage rate. For the strongly heteroscedastic data, once again the homoscedastic model performs very poorly.  Compared to the heteroscedastic model, the SUBSET prior yields a 14\% reduction in MSE and a 14\% reduction in average CI width, at a cost of a 0.014 reduction in coverage rate.}

In short, the SUBSET priors helped improve the fit of the heteroscedastic model when incorrectly specified, \green{and did much better in terms of MSE and CI widths when the true parameters lied outside of the subspace, at the cost of a small reduction in coverage rates.}

\begin{table}
    \centering
    \green{
    \begin{tabular}{crrr}
          & Homoscedastic & Heteroscedastic & SUBSET  \\
          \hline
          & \multicolumn{3}{c}{Homoscedastic data} \\
          CI Width & 1.118 & 3.111 & 1.926 \\
          CI Coverage &  0.920 & 0.943 & 0.970 \\
          MSE & 0.103 & 0.649 & 0.221 \\
          & \multicolumn{3}{c}{Mildly heteroscedastic data} \\
          CI Width & 1.363 & 3.768 & 2.685 \\
          CI Coverage & 0.378 & 0.875 & 0.806 \\
          MSE & 1.202 & 1.001 & 0.738 \\
          & \multicolumn{3}{c}{Strongly heteroscedastic data} \\
          CI Width & 3.066 & 8.330 & 7.202 \\
          CI Coverage &  0.262 & 0.871 & 0.857 \\
          MSE & 12.445 & 5.193 & 4.445 \\
    \end{tabular}
    }
  \caption{1-way ANOVA simulation study results for $\{\sigma^2_g\}_{g=1}^6$. \green{All credible intervals were at 95\%.}}
    \label{tbl:simstudy_anova}
\end{table}

\subsection{Ordinal factor covariate}
Our second simulation study used a regression setting with a single ordinal factor covariate with 9 levels.  The true group means were $0.005\times(0,1,2^4,3^4,\ldots,8^4)$.  There were five observations per group, and the residual standard deviation was 1.

We fit \green{three} models to each simulated dataset.  The first used a Zellner's $g$ prior, setting $g$ to be equal to the sample size (45).  The remaining two models were SUBSET priors using the Zellner's $g$ prior as the base prior and the following two linear subspaces:
\begin{itemize}
    \item \textit{Power:} $\mbox{span}
        \left(
            \begin{pmatrix}
                1  & 1 & \cdots & 1\\
                1 & 2^\phi & \cdots & 9^\phi
            \end{pmatrix}'
        \right)$
    \item \textit{Geometric:} $\mbox{span}
        \left(
            \begin{pmatrix}
                1 & 1 & \cdots & 1 \\
                1/\phi & 1/\phi^2 & \cdots &1/\phi^9
            \end{pmatrix}    
        \right)$
\end{itemize}
Note that the true parameter vector lies in neither of these linear subspaces.  We estimated $\phi$ alongside the true regression coefficients via Algorithm \ref{alg:is_gibbs}.  To implement this, for the \textit{power} subspace, we used 15 quantiles coming from a gamma distribution with shape equal to 2 and rate equal to 1; for the \textit{geometric} subspace, we used 15 quantiles coming from a Beta distribution with both shape parameters equal to 2. \green{To determine $\nu$, we temporarily set $\phi$ to be fixed at its mode (1 for the power subspace, and $1/2$ for the geometric subspace) and selected $\nu$ by maximizing the Bayes factor via Algorithm \ref{alg:is_fixed}.} As before, each model used 50000 posterior draws, and we generated 2000 datasets.

Table \ref{tbl:simstudy_ordcovar} provides the results for the estimation of the regression coefficients in terms of \green{95\%} CI widths, \green{95\%} CI coverage rates, and MSE.  All methods achieved \green{near 100\% coverage rates, although the different methods achieved this with varying CI widths. Compared to posterior inference using the base prior, the SUBSET priors yielded on average 9\% and 3\% smaller CI widths for the power and geometric subspaces respectively, and 25\% and 8\% lower MSE respectively.}

\begin{table}[ht]
    \centering
    \green{
    \begin{tabular}{rrrr}
        & & \multicolumn{2}{c}{SUBSET} \\
        & Zellner & (power) & (geometric) \\
        \hline
        CI Width & 3.676 & 3.337 & 3.571 \\
        CI Coverage & 0.997 & 0.997 & 0.997 \\
        MSE & 0.398 & 0.300 & 0.367 \\
        \hline
    \end{tabular}
    }
    \caption{Ordinal covariate simulation study results. \green{All credible intervals were at 95\%.}}
    \label{tbl:simstudy_ordcovar}
\end{table}

\section{Pedagogical analyses}
\label{sec:rda}
The following analyses illustrate the application of SUBSET priors using the R package \verb!SUBSET!\footnote{ 
In R, run \verb+remotes::install_github("dksewell/SUBSET") +} developed by the author.  Code to replicate these analyses is included in the Supplementary Material \citep{sewell2023supplementLSS}.

\subsection{Antihypertensive Clinical Trial (Ordinal Covariates)}
\label{subsec:antihyp}
\cite{sung2022efficacy} conducted a 7 arm randomized clinical trial aimed at discovering the effect of various drug treatments at various doses on reducing the mean sitting systolic blood pressure (MSSBP) over a period of 8 weeks.  The efficacy endpoint was the change in MSSBP from baseline to the end of the 8 week period.  For the purposes of this analysis, we will focus on the following four treatment arms: placebo, and combination treatments of telmisartan/amlodipine/chlorthalidon at quarter-dose, third-dose, and half-dose. 

Table \ref{tab:antihyp} provides the summary statistics from the study, from which it can be seen that there is a failure to observe the expected dose-response relationship between the dose of the combination treatment and the sample mean change in MSSBP (lower is better, reflecting a larger reduction in blood pressure).  However, such a non-monotonic relationship is highly implausible.  Letting $\*\mu = (\mu_1,\mu_2,\mu_3,\mu_4)$ denote the mean change in MSSBP, we can impose our prior beliefs in a monotonic dose-response relationship by shrinking our prior on $\*\mu$ to favor values on or near the span of 
\[
\begin{pmatrix}
1 & 0 \\
1 & \frac1{4^{\phi}} \\
1 & \frac1{3^\phi} \\
1 & \frac1{2^\phi} 
\end{pmatrix}
\]
for some $\phi>0$. That is, we believe that there may be some placebo effect, and on top of that there is a drug effect that follows some power law.

\begin{table}
    \centering
    \begin{tabular}{c|rr}
        & Mean & SD \\
        \hline
        Placebo & -5.85 & 10.74 \\
        $1/4$ dose & -18.87 & 16.87 \\ 
        $1/3$ dose & -14.55 & 14.84 \\
        $1/2$ dose & -19.55 & 14.75
    \end{tabular}
    \caption{Summary statistics from multi-arm clinical trial on antihypertensive drugs.}
    \label{tab:antihyp}
\end{table}

As a base prior, we used independent conjugate Normal-Gamma distributions over the mean and precision for each treatment arm.  Hence our model for the antihypertensive randomized clinical trial is
\begin{samepage}
\begin{align} \nonumber
y_i|\mbox{treatment}_i=k,\boldsymbol{\mu},\boldsymbol{\tau} &\overset{ind}{\sim} N(\mu_k,\tau_k),& \\  \nonumber
\mu_k|\boldsymbol{\tau} &\overset{ind}{\sim} N(a_0, b_0\tau_k), &\\
\tau_k &\overset{ind}{\sim} \Gamma(c_0/2,d_0/2),&
\end{align}
\end{samepage}
for $k=1,\ldots,4$, where we set $a_0=-5$, $b_0 = 1$, $c_0=3$, and $d_0=75$.  Additionally, we set a $\Gamma(2,2)$ prior on $\phi$, with a prior mean value of $\phi$ equal to 1 (linear drug effect). \green{We obtained 10000 draws each from the posterior and prior distributions under the base prior and the posterior under the SUBSET prior.}

\begin{figure}
    \begin{center}
        \includegraphics[width = 0.75\textwidth]{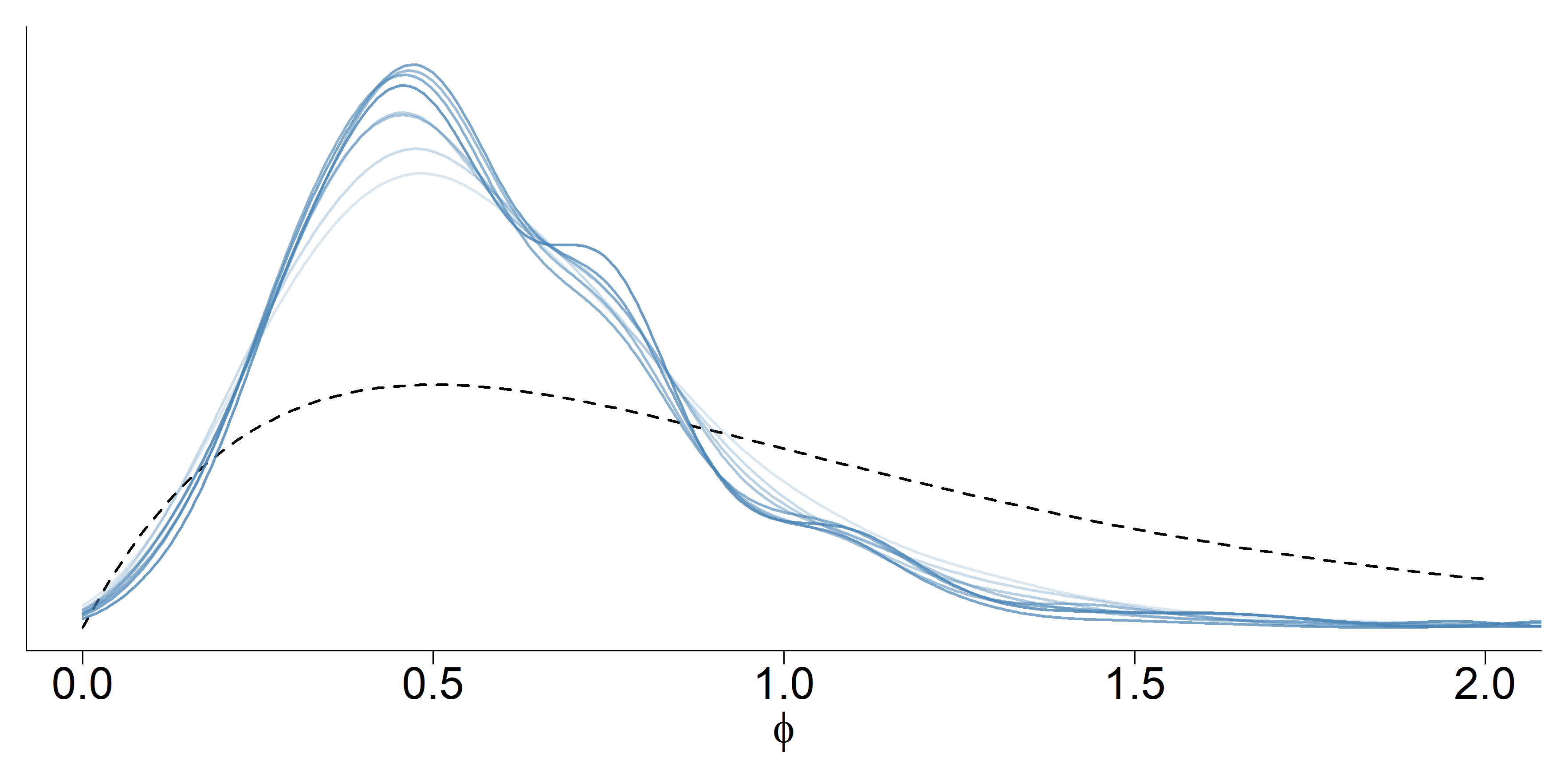}
    \end{center}
    \caption{\green{Antihypertensive clinical trial example.  Dashed line shows the $\Gamma(2,2)$ prior on $\phi$ (the parameter dictating the power law of the treatment effect), and the solid lines correspond to the posterior of $\phi$ using values of $\nu$ ranging from 0.25 to 2 graded from lightest to darkest respectively.}}
    \label{fig:antyihyp-phi}
\end{figure}

We ran the Gibbs sampler of Algorithm \ref{alg:is_gibbs} (similar results using Algorithm \ref{alg:gibbs_spline} are provided in the Supplementary Material) using values of $\nu$ in $(0.25,0.5,\ldots,\green{2})$ and for $\phi$ considered \green{50} evenly spaced quantiles from its Gamma prior.  Figure \ref{fig:antyihyp-phi} shows that for most values of $\nu$, the posterior of $\phi$ centers near \green{$1/2$}, implying a prior that shrinks towards a \green{square} root relationship between dose and mean change in MSSBP.

\begin{figure}[h]
    \begin{center}
        \includegraphics[width = 0.75\textwidth]{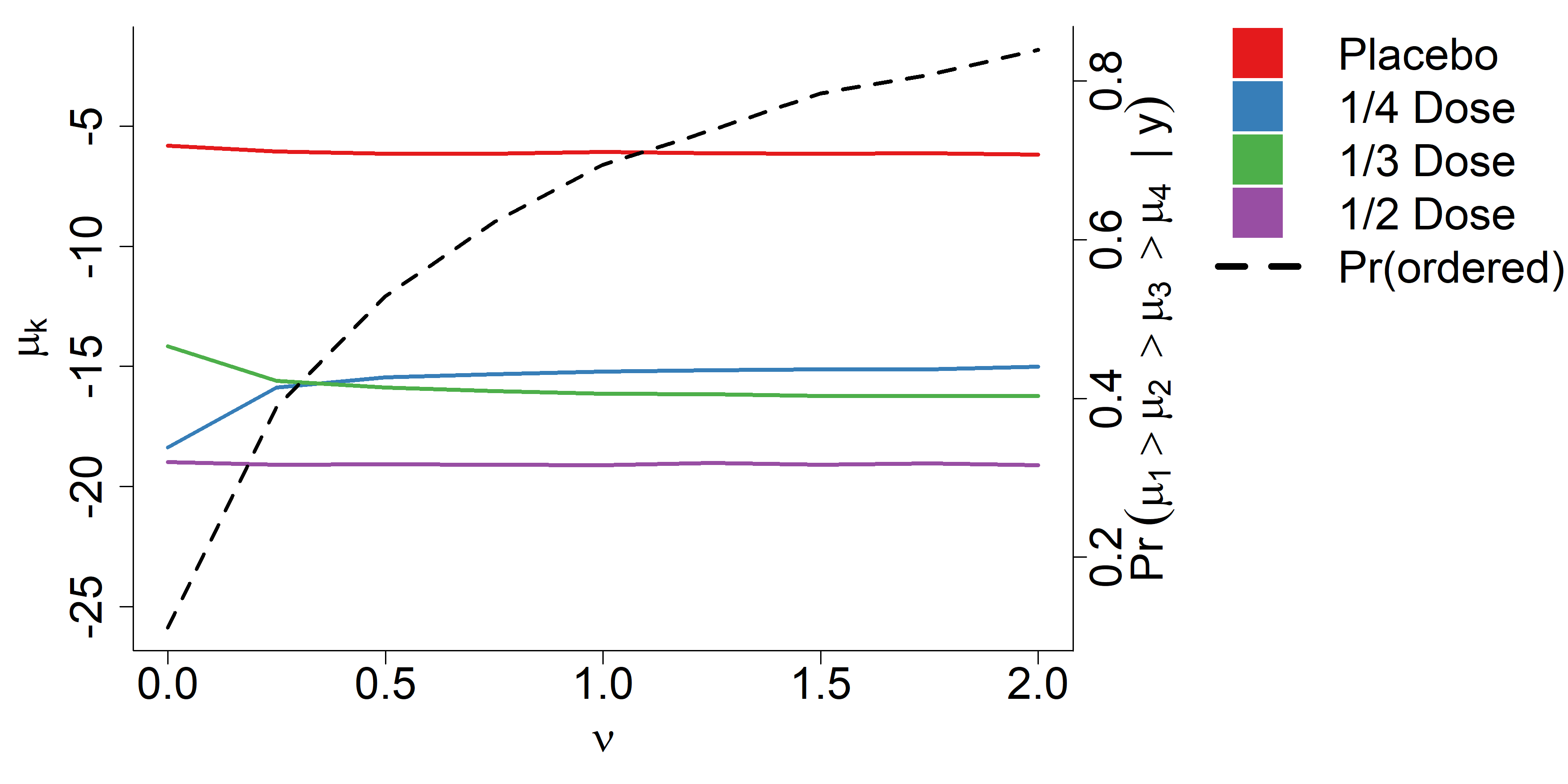}
    \end{center}
    \caption{\green{Antihypertensive clinical trial example.  The horizontal axis represents differing levels of $\nu$, i.e., shrinkage towards the linear subspace; the left vertical axis corresponding to the solid lines shows the posterior mean estimate of the mean change in MSSBP; the right vertical axis corresponding to the dashed line shows the posterior probability that there is a monotonic relationship between dose and mean change in MSSBP. As $\nu$ increases, the expected dose-response relationship emerges.}}
    \label{fig:antihyp-transition}
\end{figure}

Figure \ref{fig:antihyp-transition} shows for the different values of shrinkage, i.e., $\nu$, the posterior mean for the mean change in MSSBP\green{, as well as the posterior probability that the dose response is monotonic}.  Although $\nu$ increases, the point and interval estimates for the placebo are near constant.  Importantly, however, for \green{$\nu\geq 0.5$}, the posterior means of $\*\mu$ reflect a dose-response relationship, i.e., $\E_{\pi_\nu}(\mu_1|y) > \E_{\pi_\nu}(\mu_2|y) > \E_{\pi_\nu}(\mu_3|y) > \E_{\pi_\nu}(\mu_4|y)$, thereby giving us plausible estimates on the effect of the combination treatment on reducing hypertension. \green{While under the base prior there was only a 0.11 posterior probability of a monotonic relationship, this surpassed 0.5 at $\nu=0.5$ and 0.8 at $\nu=1.75$.}

\subsection{Influenza and Pneumonia Monthly Mortality (Smoothing MA($q$) coefficients)}
\label{subsec:pneumonia}
As a second example, we illustrate how to use SUBSET priors to smooth the estimates of a sequence of parameters.  We analyzed monthly mortality caused by influenza and pneumonia in the US from 2014-2019 \citep{nchs2022allcausemortality}.  We detrended the data and based on ACF and PACF plots fit a \green{moving average (MA) model with 15 lags} using \green{adaptive MCMC \citep{scheidegger2021adaptMCMC}}.  We used as the base prior for the MA coefficients $N(0,2)$, and for the \green{variance of the residuals a gamma distribution with shape and rate both equal to 2.}  \green{We obtained 50000 posterior draws under $\pi_0$ and removed 10000 as a burn-in period, for a remaining 40000 draws.  We obtained 5000 draws from the prior to estimate the Bayes factors.}

To shrink towards smoothed estimates of the MA coefficients, we used Algorithm \green{\ref{alg:is_fixed}} using the linear subspace spanned by the natural cubic spline basis functions evaluated at the lags (1-15) using \green{4} degrees of freedom (implying \green{3} internal knots).  \green{The value of $\nu$ which maximized the Bayes factor was 32.7.}

\begin{figure}
    \begin{center}
       \includegraphics[width = 0.75\textwidth]{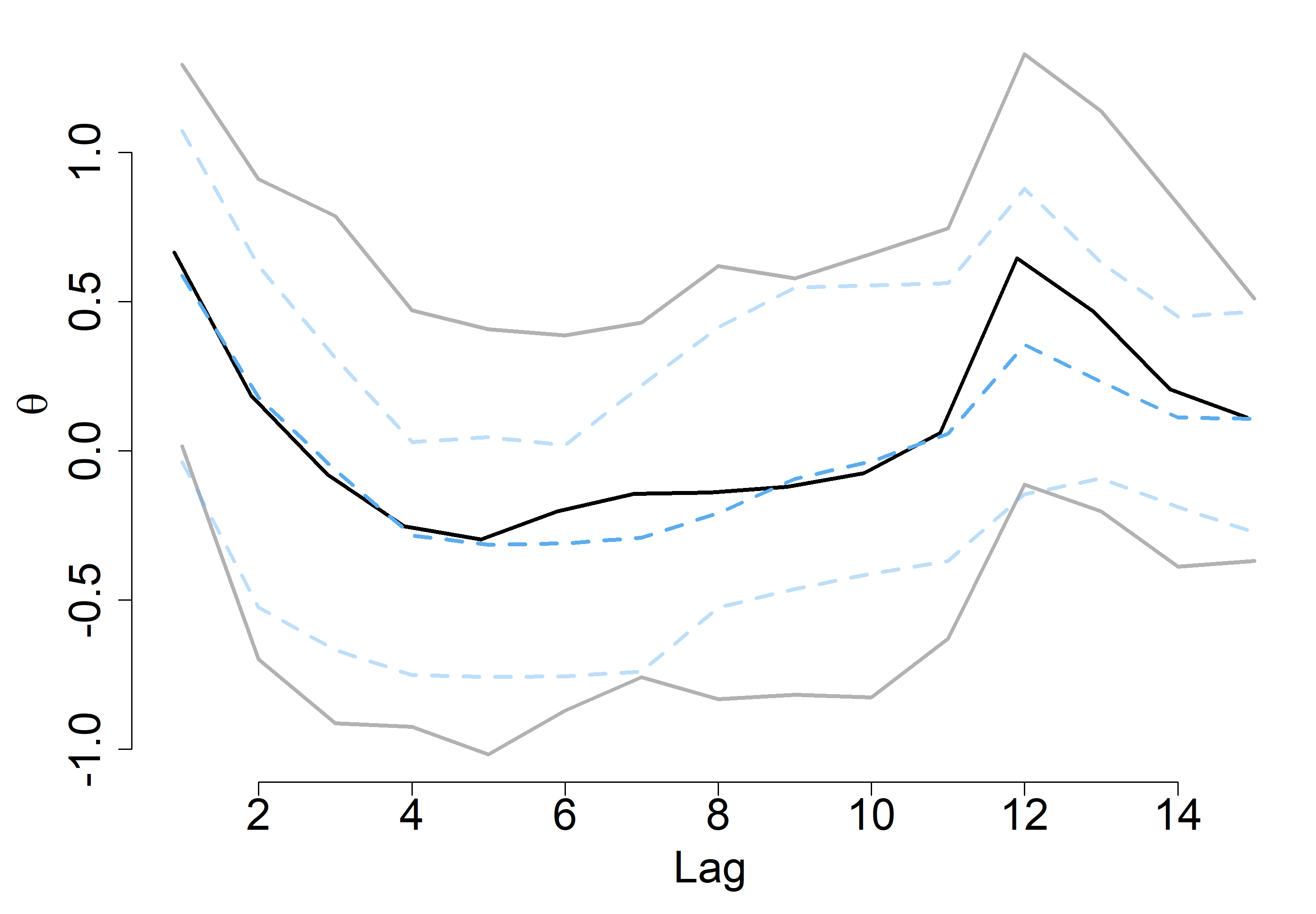}
    \end{center}
    \caption{\green{Influenza and pneumonia monthly mortality example. Estimated MA coefficients. Posterior mean and 95\% credible interval bounds under the base prior are given by the solid black and gray lines respectively. Under the SUBSET prior with $\nu=32.7$ (selected via Bayes factor), these are given by the dark blue and light blue dashed lines respectively.}}
    \label{fig:flu-MA}
\end{figure}

Figure \ref{fig:flu-MA} shows the posterior mean and 95\% credible intervals for the MA coefficients under both the base prior and the SUBSET prior with $\nu=\green{32.7}$.  From this figure we can see that both posteriors are telling the same overall story, but that the posterior point and interval estimates under the SUBSET prior are much more smooth.

\section{Discussion}
\label{sec:conclusion}
The information and beliefs we have about parameters of interest are often relational in nature, which cannot be encoded simply by, e.g., a location shift in the prior distribution.  Instead, such prior knowledge leads us to believe that the parameters ought to lie on or near some linear subspace. This type of information is ubiquitous, and yet a comparably small amount of attention has been given it. Previous work has focused almost exclusively on point estimation within a regression setting.

We have proposed a new approach to incorporating relational prior information that can be represented by parameters lying on a linear subspace.  Our approach has the following advantages.  First, our approach is completely generalizable to any setting, including regression. 

Second, we argue that the most logical way to handle prior beliefs described by a linear subspace is to incorporate such information in the prior distribution, which by definition is where our prior beliefs ought to be contained.  Doing so allows not only point estimates but all inference to account for this information. 

Third, our approach of applying exponential tilting to a base prior does not ``overwrite'' other prior information that has already been encoded in this base prior such as previous data or scientific domain knowledge.  Rather, our approach takes the a priori plausible regions of the parameter space from this prior information and further hones the plausible regions to conform to prior beliefs described by the linear subspace.

Fourth, we have provided methods to obtain posterior inference in a highly computationally efficient manner, allowing for researchers to quickly \green{derive Bayes factors for or conduct sensitivity studies of $\nu$ or other facets of the linear subspace. In particular, once $K_0$ and $K_y$ prior and posterior samples respectively have been obtained under the base prior, computing the Bayes factor in Algorithm \ref{alg:is_fixed} for a particular $\nu$ costs $\~O(K_0\vee K_y)$. Performing the MH-within-Gibbs sampler of Algorithm \ref{alg:is_gibbs} has a bottleneck from the multinomial importance resampling of the $K_y$ posterior samples at each step of the Gibbs sampler, leading to a computational cost of $\~O(K_\nu K_y \log (K_y))$ \citep{kronmal1979alias}, assuming that $K_0$ grows at the same or slower rate as $K_y$.  
Algorithm \ref{alg:gibbs_spline} alleviates this problem entirely by avoiding the resampling of the original $K_y$ posterior draws, and has a computational cost of $\~O(K_0\vee K_\nu)$. Critically, these computational costs are free from the sample size and the computational complexity of evaluating the posterior. 
}

Our proposed methodology has certain limitations.  First, \green{while the use of Bayes factors provides an automated approach to selecting the hyperparameter $\nu$ for the case of fixed $\phi$, for the case when $\phi$ is estimated, $\nu$ may be determined again using Bayes factors for a user-specified value of $\phi$ else, as an anonymous reviewer pointed out, one may consider using an alternative approach such as cross-validation or putting a prior on $\nu$ and estimating it.}  Further research into the selection of $\nu$ in this context would be worthwhile.  Second, with unknown $\phi$, the estimation of the normalizing constant $Z_{\nu,\phi}$ may become challenging with higher dimensional $\phi$.  Third, in the proposed estimation algorithms, the normalizing constant $Z_{\nu,\phi}$ is estimated at least once through a Monte Carlo approach, and in Algorithm \ref{alg:gibbs_spline} a second time through splines.  While potentially concerning, in our simulation study (see Supplemental Material for results), this did not seem to have deleterious effects on estimation and inference, and should it appear to be problematic in new scenarios not considered here, work on doubly intractable posteriors \citep[see, e.g.,][]{park2018bayesian} may be brought to bear. \green{Fourth, Algorithm \ref{alg:is_gibbs} relies on using a discrete number of values of $\phi$.  We anticipate that this will not typically be problematic in practice since $\phi$ will not be a parameter of primary interest, yet there may be situations where a discretization of $\Phi$ is suboptimal.}

It is not uncommon for data scientists to obtain unexpected, and perhaps unreasonable, results, such as in the antihypertensive clinical trial of Section \ref{subsec:antihyp} in which the quarter-dose yielded a greater reduction in MSSBP than the third-dose or the half-dose.  A reaction such as ``this can't be right!'' is indicative that there are in fact prior beliefs about the relations between the model parameters. In the above example, we should feel extremely confident that the reduction from a quarter-dose ought to be less than or equal to that from a third-dose, which in turn ought to be less than or equal to that of a half-dose.  This type of prior belief also appears when there is an ordinal covariate in a regression setting. Other examples include when we expect smoothness (of which the monotonicity above was a special case) across a naturally ordered set of parameters, when we believe that there may be near homoscedasticity, or when we believe there may be equal response rates in a two-population binomial setting. Our proposed sampling algorithms outlined in Section \ref{sec:estimation} should be reasonably easy to implement for a practicing statistician accustomed to performing Bayesian analyses; still, to further lower the barrier to implementation we have developed the R package \verb!SUBSET! and have illustrated its use in the Supplementary Material where the real data analyses are replicated.

\FloatBarrier

\begin{supplement}
\stitle{Supplementary Material to ``Posterior shrinkage towards linear subspaces''}
\sdescription{This supplementary material provides the proofs for Theorems 2-3, the derivation for the Bayes factors, simulation study results for the large sample approximations, as well as a walkthrough of how to implement the R package SUBSET, including replication of the findings presented in Section \ref{sec:rda}.}
\end{supplement}

\bibliographystyle{ba}
\bibliography{SewellBibFiles}


\end{document}